# How can we think the complex?

Carlos Gershenson and Francis Heylighen
Centrum Leo Apostel, Vrije Universiteit Brussel
Krijgskundestraat 33. B-1160 Brussels, Belgium
http://www.vub.ac.be/CLEA , {cgershen,fheyligh}@vub.ac.be

This chapter does not deal with specific tools and techniques for managing complex systems, but proposes some basic concepts that help us to think and speak about complexity. We review classical thinking and its intrinsic drawbacks when dealing with complexity. We then show how complexity forces us to take build models with indeterminacy and unpredictability. However, we can still deal with the problems created in this way by being adaptive, and profiting from a complex system's capability for self-organization, and the distributed intelligence this may produce.

## 1. Classical Thinking

The majority of scientific models—as well as much of our intuitive understanding—implicitly rely on a "classical" or Cartesian mode of thinking, which is expressed most explicitly in the classical or Newtonian mechanics that dominated the scientific worldview until the beginning of the 20th century. It is based on the following assumptions (Heylighen, 1990):

- *reductionism* or *analysis*: to fully understand a system you should decompose it into its constituent elements and their fundamental properties.

- *determinism*: every change can be represented as a trajectory of the system through (state) space, i.e. a linear sequence of states, following fixed laws of nature. These laws completely determine the trajectory towards the future (predictability) as well as towards the past (reversibility).

- *dualism*: the ultimate constituents of any system are particles, i.e. structureless pieces of *matter* (*materialism*). Since matter is already completely determined by mechanistic laws, leaving no freedom for intervention or interpretation, the only way we can include human agency in the theory is by introducing the independent category of *mind*

- *correspondence theory of knowledge*: through observation, an agent can in principle gather complete knowledge about any system, creating an internal representation whose components correspond to the components of the external system. This establishes a single, true, objective mapping from the realm of matter (the system) to the realm of mind (the representation).

- *rationality*: given such complete knowledge, in its interaction with the system an agent will always choose the option that maximizes its *utility* function. Thus, the actions of mind become as determined or predictable as the movements of matter.

These different assumptions are summarized by the principle of *distinction conservation* (Heylighen, 1989, 1990): classical science begins by making as precise as possible distinctions between the different components, properties and states of the system under observation. These distinctions are assumed to be absolute and objective, i.e. the same for all observers. They follow the principles of Aristotelian logic: a phenomenon belongs either to category **A**, or to **not A**. It cannot be both, neither, in between, or "it depends". The evolution of the system conserves all these distinctions, as distinct initial states are necessarily mapped onto distinct subsequent states, and vice-versa (causality, see (Heylighen, 1989)). Knowledge is nothing more than another such distinction-conserving mapping from object to subject, while action is a mapping back from subject to object.

Of course, we know that these assumptions represent ideal cases that are never realized in practice. Yet, most educated people still tend to assume that a complete, deterministic theory is an ideal worth striving for, and that the scientific method will lead us inexorably to an ever closer approximation of such objective knowledge. The lessons from complexity research point in a different direction, though.

## 2. Complexity

What is complexity ? Let us go back to the Latin root *complexus*, which means "entwined" or "embraced". This can be interpreted in the following way: in order to have a complex you need: 1) two or more distinct parts, 2) that are joined in such a way that it is difficult to separate them. Here we find the basic duality between parts which are at the same time *distinct* and *connected*. Therefore, the analytical method alone won't allow us to understand a complex, as by taking apart the components it will destroy their connections.

This means that using classical methods the behaviour of a complex system will be hard to describe, explaining the connotation of "difficult" that the word "complex" has later acquired. The components are mutually entangled, so that a change in one component will propagate through a tissue of interactions to other components which in turn will affect even further components, including the one that initially started the process. This makes the global behaviour of the system very hard to track in terms of its elements. Unlike the simple, "billiard-ball-like" systems studied by classical mechanics, complex systems are the rule rather than the exception. Typical examples are a living cell, a society, an economy, an ecosystem, the Internet, the weather, a brain, and a city. These all consist of numerous elements whose interactions produce a global behavior that cannot be reduced to the behavior of the separate coompontent.

Complexity is itself a complex concept, as we cannot make a unambiguous distinction between simple and complex systems. Many measures of complexity have been proposed for different contexts, such as computational, social, economic, biological, etc. (Edmonds, 2000).

However, there is no universal measure that would allow us to establish the degree of complexity of an arbitrary system. Yet, within an agreed frame of reference, we can sometimes compare two systems, noting that the one is more complex than the other. Thus, complexity at best determines a partial ordering, not a quantitative measure. Overall, we can say that *the complexity of a system increases with the number of distinct components, the number of connections between them, the complexities of the components, and the complexities of the connections*. This is a recursive definition that is general enough to be applied in different contexts. For example, everything else being equal, a firm will be more complex than another one if it has more divisions, if its divisions have more employees, if the divisions have more channels of interaction, and/or if its channel of interactions involve more person-to-person interactions.

While we do not really need an absolute measure of complexity, using such relative comparison may be useful to indicate when it becomes necessary to abandon our simple, classical assumptions and try to develop a more sophisticated model. Shifting from classical to "complex" thinking brings both gains and losses. Let us start with the expectations we have to abandon, and then point out some of the novel insights we gain.

### 3. Indeterminacy

Relinquishing classical thinking means giving up the principle of distinction conservation. This implies, first, that we can no longer assume given, invariant distinctions: a distinction made by one observer in one context may no longer be meaningful—or even possible—for another observer or in another context.

This point was made most forcefully in quantum mechanics (Heylighen, 1990): in some circumstances, an electron appears like a particle, in others like a wave. Yet, according to classical thinking, particle and wave are mutually exclusive categories. In quantum mechanics, on the other hand, the "particle" and "wave" aspects are *complementary*: they are jointly necessary to characterize the electron, but they can never be seen together, since the observation set-up necessary to distinguish "particle-like" properties is incompatible with the one for "wave-like" properties. This was formulated by Heisenberg as the principle of *indeterminacy:* the more precisely we distinguish the particle-like properties, the more uncertain or indeterminate the wave-like properties become.

A more intuitive example of indeterminacy is the well-known ambiguous figure that sometimes looks like a rabbit, sometimes like a duck. While both "gestalts" are equally recognizable in the drawing, our perception—like a quantum observation set-up—is incapable to see them simultaneously, and thus tends to switch back and forth between the two interpretations. Complementary properties, like the rabbit and duck gestalts, are distinct yet joined together. But while we see the one, we cannot see the other!

Because of the correspondence assumption, classical thinking tends to confuse what things are and how we see or know them to be. Thus, observers have engaged in controversies on "what things are", while actually disagreeing on how to model or represent these phenomena. When we speak about a phenomenon it is hard to specify whether we refer to the representation or to the represented, because our language does not make such a distinction, using the verb "to be" for both. To avoid such confusion we have proposed an ontological distinction between "absolute being" and "relative being" (Gershenson, 2002). The *absolute being* (abs-being) refers to what the thing actually is, independently of the observer (Kant's "Ding-an-sich"). The *relative being* (rel-being) refers to the properties of the thing as distinguished by an observer within a context. Since the observer is finite and cannot gather complete information, rel-beings are limited, whereas abs-beings have an unlimited number of features. Therefore, there exists an infinity of potential rel-beings for any abs-being.

We can illustrate this abstract notion by imagining a sphere which is black on one hemisphere and white on the other, as depicted in Fig. 1. Suppose we can observe the sphere only from one perspective. For some, the sphere will (rel)be white, for others it will (rel)be black, for others it will (rel)be half black and half white, and so on. How can we decide of which colour the sphere (abs) is? Taking an average does not suffice, since it could be the case that more than ninety percent of people see the sphere white, and we would conclude that it *is* mostly white, while it actually (abs)is half white and half black. The best we can do is to indicate the perspective (context) for which the sphere (rel)is of a particular colour. With real systems, we will never reach their abs-being, because there are always more properties (dimensions) than we are aware of. This task would be like determining the colour of an infinite-dimensional sphere when you can only see one two-dimensional projection at a time.

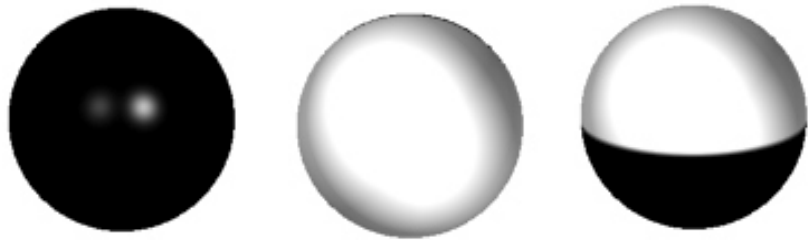

Figure 1: the same black-and-white sphere seen from three different angles

With simple systems such as the 3-dimensional sphere, the number of rel-beings is limited. However, complex systems have so many types of components and connections that observers can have rel-beings that are so different that it may appear impossible to recognize them as aspects of the same thing. For example, organizations have been described using metaphors such as an organism, a machine, a brain, a community, a market, and a political power game, and models such as a hierarchy, a network, and a linear input-output system. For a classically thinking executive, this constant shift in models and concomitant management styles is bewildering, as it seems that only one (or none) of these approaches can be correct. Yet, an organization has both mechanistic and organic aspects, is simultaneously a cooperative community and a competitive arena, a rule-bound system and an open, creative environment, a hierarchy and a network...

There is no "best" model, as different rel-beings are appropriate for different contexts, and different purposes (Beer, 1966; Heylighen, 1990). With a classical way of thinking, we can spend all our efforts in trying to decide what *is* the system. Complex thinking, on the other hand, allows us to contemplate different representations at the same time (e.g. by proposing a *metarepresentation* (Heylighen, 1990)), in order to have a less-incomplete understanding of the system. To tackle concrete problems, we can then choose the representation that is most appropriate for that specific context, being well aware that a different problem may require a radical shift in representation. For example, when tackling internal conflicts it may be useful to see a firm as a network of interdependent communities; when optimizing production, as a matter- and information-processing mechanism.

### 4. Non-linearity and chaos

According to classical thinking, distinctions are invariant not only over observers, but over time. The principle of causality can be formulated as "equal causes have equal effects", or equivalently as "effects co-vary with their causes". This is nothing more than a statement that the distinctions between causes or initial states must necessarily carry through to their effects, and vice-versa. While we may hold this principle to be true at the level of absolute being, i.e. the complete things-in-themselves (microscopic causality, Heylighen, 1989), it is in general not true at the level of relative being, i.e. the coarse, finite distinctions made by an observer. Thus, microscopic causality (i.e. determinism) does not in itself produce macroscopic causality (i.e. predictability). This can be inferred most directly from the existence of (deterministic) *chaos*, which follows from the *non-linearity* that characterizes complex systems.

A system is *linear* if effects (outputs) are proportional to their causes (inputs). For example, if you put twice as much ore in your furnaces, the plant will produce roughly twice as much steel. This can be understood through the principle of conservation of energy and matter: the amount that comes out depends directly on the amount you put in (though there will of course be a few losses here and there). But what happens if (part of) the output is redirected and added back to the input? In principle, the next output will be larger, since it uses both the input and the previous output, and therefore no longer proportional to the input alone. The next output will be even larger, as it uses not only the new input but the two previous outputs. For example, a firm can reinvest some of the money it gets for its products to increase production. Increasing production brings in more money and thus further increases production, leading to an exponential growth in output.

Thus, non-linearity can be understood as the effect of a causal loop, where effects or outputs are fed back into the causes or inputs of the process. Complex systems are characterized by networks of such causal loops. In a complex, the interdependencies are such that a component A will affect a component B, but B will in general also affect A, directly or indirectly. A single feedback loop can be positive or negative. A positive feedback will amplify any variation in A, making it grow exponentially. The result is that the tiniest, microscopic difference between initial states can grow into an macroscopically observable distinction.

This is called *sensitive dependence on initial conditions*, and is a defining feature of chaos. Because the initial difference is too small to perceive, the principle of causality cannot help us in predicting the final outcome. A well-known example of such a difficult-to-predict, chaotic system is the weather, as the fluttering of a butterfly in Tokyo can grow into a hurricane devastating New York. The observation that small causes can have large effects is obvious in social systems as well. For example, during a tense negotiation the tiniest hint of a smile on the lips of a CEO may create the impression with the other party that this guy is not to be trusted, thus leading them to harden their stance, and finally reject a billion-dollar merger operation. Such a system in a sense *creates* distinctions, as an indistinguishably small difference in initial state leads to macroscopically distinct outcomes.

The inverse of the amplifying effect of positive feedback is the dampening effect of negative feedback. Here any variation is counter-acted or resisted, bringing the system back to its equilibrium state. As a result large causes (variations) may have little or no effect. For example, an entrenched culture in an organization can be very difficult to change, as new measures are actively or passively resisted, ignored or deflected. Such a system destroys distinctions, as distinct causes lead to the same outcome.

Complex systems will typically exhibit a tangle of interconnected positive and negative feedback loops, where the effects of any change in a component cascade through an increasing number of connected components, in part feeding back, positively and/or negatively, into the initial component. If there is a variable time delay between these effects, it becomes in principle impossible to make predictions, because we do not know who will affect who first and thus whether an effect will be dampened before it has had the chance to get amplified or not (cf. Gershenson *et al.*, 2003). An example can be found in the stock exchange where stocks are bought and sold depending on their price, while the price is determined by how much is bought and sold. This intrinsic feedback loop has both negative and positive aspects. The law of supply and demand implies a negative feedback, since an increase in price normally reduces the demand, and this—after a variable delay—will reduce the price again. However, the parallel mechanism of speculation entails a positive feedback, as an increasing price makes buyers anticipate an even higher price in the future, thus enticing them to buy more of the stock now. The interaction between both non-linear effects produces the chaotic movement of stock prices that we know so well.

In the simpler situation where the delays are known (or can be neglected), it is sometimes possible to get at least a qualitative estimate of what can happen by identifying the signs (positive or negative) and the strengths of the different feedback loops in the network of influences. This method, which is ofter used to build computer simulations, is developed in the discipline of system dynamics (Sterman, 2000).

## 5. Adapting to Complexity

Given the intrinsic unpredictability of complex systems, how can we design, build or generally deal with them? First we have to accept that we will never able to control or predict their behavior completely. It is only natural that there will be surprises, errors and problems, as there have always been. However, we can always try to cope with unexpected events by adapting our actions to the new situation—if necessary reconfiguring the system without destroying it. Different principles and methods for a*daptation* have been investigated in cybernetics (Heylighen and Joslyn, 2001), artificial intelligence (Russell and Norvig, 1995), neural networks (Rumelhart et al., 1986), multi-agent systems (Wooldridge, 2002; Schweitzer, 2003), genetic algorithms (Mitchell, 1998), chaos control (Chen and Yu, 2003), and many other disciplines. Research is going on still, trying to design and build systems that are even more adaptive.

To adapt to any change, whether anticipated or not, it suffices to compensate for any perceived deviation of the actual situation from the desired course. This is the basic method of *feedback control:* correcting errors after the fact (Heylighen & Joslyn, 2001). If the reaction comes quickly enough, before the problem has had the chance to grow, feedback-based regulation can be extremely effective. The core innovation that engendered the field of cybernetics was the observation that it does not matter how complicated the system of factors and interactions that affect a variable that we wish to keep under control: as long we have some means of counteracting the deviation, the underlying causality is irrelevant (Kelly, 1994). For example, it does not matter which complicated combination of social, political or technological changes causes an economy to overheat: in general, the central bank can regulate the rate of growth simply by increasing its interest rates.

Feedback control, however, still requires that we have a sufficiently broad repertoire of counteractions at our disposal (*requisite variety*), and that we know which action to execute in which circumstances (*requisite knowledge*) (Heylighen & Joslyn, 2001). The cybernetic law of requisite variety notes that the greater the variety of perturbations that the system may be subjected to, the larger the variety of actions it needs to remain in control. Since the number of perturbations that a present-day organization may encounter is in practice unlimited, the recommendation is for that organization to maximize the diversity of its possibilities for intervention, as you can never predict what will be necessary when.

However, in order to react quickly and appropriately, it is good to have at least an *expectation* of what may happen and which reaction would be appropriate. Expectations are subjective probabilities that we learn from *experience*: the more often circumstance B appears after circumstance A, or the more successful action B is in solving problem A, the stronger the association A → B becomes. The next time we encounter A (or a circumstance similar to A), we will be prepared, and more likely to react adequately. The simple ordering of options according to the probability that they would be relevant immensely decreases the complexity of decision-making (Heylighen, 1994). Thus, we do not need deterministic models or predictions: having realistic default expectations with the possibility to correct for errors or exceptions after they have appeared works pretty well in practice. Moreover, we can tackle as yet unencountered combinations of circumstances by aggregating the recommendations made by different learned associations in proportion to their strength, using the method of *spreading activation* (Heylighen, 1999).

That is how our brains deal everyday with other complex systems: colleagues, children, pets, computers, etc. However, much to our dismay and frustration, most designed systems still lack this characteristic. For example, computers programmed according to rigid rules cannot recover on their own when something goes wrong (Heylighen & Gershenson, 2003). Organizations rarely have procedures to learn from experience, and those that do usually try to store acquired knowledge as formal data rather than as a constantly evolving network of associations between circumstances and actions. The only working paradigm to do this outside the brain are neural network simulations, but these are usually limited to very specialized applications such as hand writing recognition, rather than broad issues of management and procedure. They are also extremely difficult to analyze. Yet, our research suggests that these "brain-like" mechanisms are easily extended to the organizational level, thus supporting an adaptive, collective intelligence (Heylighen, 1999).

## 6. Self-organization

While adaptation tells us how we may cope with complexity, we can also have the complex systems work for us. Indeed, any dynamical system is in principle capable of *self-organization* (Ashby, 1962; Heylighen, 2002; Heylighen & Gershenson, 2003), and the more complex the system, the more "interesting" the results. We can define *organization* as *structure* with *function*. Our definition of complexity already implies structure, which we see as the combination of distinction (differentiation) and connection (integration). Function means that the structure is developed to achieve some goal or purpose. The basic mechanism of self-organization is that sooner or later dynamic systems evolve to an *attractor* of the dynamics, i.e. a stable configuration of states, which the system can enter but not leave. We can say that the components in this configuration have mutually adapted (Ashby, 1962), limiting their interactions to those that allow this collective configuration to endure.

Thus, the "function" of this configuration is to survive, and the further behavior of the system can be understood as supporting that function. For example, a firm will sooner or later evolve its own culture consisting of a range of unwritten rules and preferences that constrain the employees' behavior. While "officially" the purpose of any rules may be to maximize the productivity of the firm, in reality the function of this culture will be basically to maintain itself. While self-maintenance implies a minimal level of productivity so that the firm does not go bankrupt, attempts to further increase the workload may be resisted because they endanger the social culture itself.

The advantage of self-organizing systems is that they search by themselves for solutions, without the need for a manager or engineer intervening. Moreover, they are intrinsically open and flexible: when unexpected changes come about, they can adapt seamlessly, without the need for centralized control. The organization is distributed over all the participating components and their connections. This means that the system is *robust*: it can survive destruction of part of its components without too much damage, as the other components make up for the lost functions.

Self-organization contradicts the classical way of thinking in several respects. To start with, self-organization by definition cannot be reduced to independent components, as it emerges wholly out of their interactions. Then, self-organization not only allows, but thrives on, randomness or indeterminacy: the more perturbations or fluctuations the system undergoes, the quicker it will reach an attractor, and the more stable the eventual attractor will be. This is the *order from noise* principle (von Foerster, 1960; Heylighen, 2002). Of course, too strong perturbations will destroy any organization, and therefore natural systems will typically evolve to the "edge of chaos" where there is sufficient variation to make the system creative and flexible, but not enough to make it wholly chaotic. Self-organization is a non-linear process, since tiny perturbations can decide which out of two completely different attractors the system enters, while once the attractor is reached, major perturbations will have little or no effect. Most fundamentally, self-organization does not conserve distinctions: it not only destroys the distinction between two initial states that end up in the same attractor, while creating a distinction when the same initial state can end up in two different attractors; it moreover creates and destroys distinctions at a higher level by integrating components in a new coherent, organizationally "closed" system (Heylighen, 1991, 2002) that has novel properties.

This is the phenomenon of *emergence*. Emergent properties characterize a system that cannot be reduced to the sum of its parts. For example, the weight of a gas is the sum of the weights of all the molecules in that gas. But the temperature and pressure of the gas simply do not exist for a single molecule, as they measure the intensity of interactions *between* the molecules. Gold is yellow, shiny, and malleable, but we cannot deduce these properties by observing the atoms of gold (Anderson, 1972). We say that a cell is alive, but it is made of non living elements.

We cannot fit emergence within a classical framework since it presupposes that things can only be one thing at a time. If a cell is a collection of molecules and molecules are not alive, then how could a cell be alive? We cannot speak about different levels of abstraction, since classical thinking assumes that there is only one "true" representation. But in complex thinking we can characterize the same abs-being at different levels using different rel-beings. Then the mystery disappears, and we can speak comfortably about emergence. We can understand it as a process that requires us to change our model of a system (Rosen, 1985; Heylighen, 1991) in order to better understand and predict the system's behavior (Shalizi, 2001).

### 7. Distributed intelligence

Self-organization even deals a blow to the dualism of classical thinking, as it blurs the distinction between matter and mind. Organization clearly cannot be reduced to the matter of its elements, as the same organization can be implemented in different material substrates (e.g. brains and computers), whereas an isolated particle of matter lacks any organization. But is organization a form of mind?

Our definition as *structure with function* does include the apparently "mental" property of goal-directedness. However, the fundamental insight of cybernetics is that goal-directedness can be understood as a type of negative feedback loop which is can implemented just as easily in mechanical systems, such as a thermostat or an automatic pilot, and in living and cognitive systems, such as the brain. What allows us to model all of these in a similar way is the concept of *information*, which has been defined by Bateson (1972) as "a difference that makes a difference". Whether this difference is carried by material objects, spoken language or electrical impulses in the brain is in se irrelevant: what counts is in how far it helps the system to understands its actual situation and take appropriate action so as to reach its goal—in spite of constantly changing circumstances. Information is what reduces the uncertainty as to the next thing that is going to happen or next action to take.

Self-organization can be seen as a spontaneous *coordination* of the interactions between the components of the system, so as to maximize their synergy. This requires the propagation and processing of information, as different components perceive different aspects of the situation, while their shared goal requires this information to be integrated. The resulting process is characterized by *distributed cognition* (Hutchins, 1995): different components participate in different ways to the overall gathering and processing of information, thus collectively solving the problems posed by any perceived deviation between the present situation and the desired situation. For example, the navigation of a large ship, as investigated in detail by Hutchins (1995), requires the activity of several people in different roles coordinated by means of formal and informal procedures, instruments, maps, ship navigation manuals and a variety of communication channels.

Some of these "distributed" components are intelligent, cognitive agents, such as human beings, but others are mere physical supports, such as notebooks, flip charts, or telephones, that store and transmit information between the agents. From the cybernetic perspective, there is no strict boundary between the "material" and the "mental" components: the information stored in my notebook is as much a part of my memory as is the information stored in my brain, and my email connection is as much part of my communication equipment as are the nerves that control my speech. What counts is in how far the information carried by each medium is *under control*, i.e. can be registered and transmitted as easily and reliably as desired. Thus, for remembering appointments I may trust my notebook more than my brain, and for explaining a complex task to a colleague I may rely more on an email message than on a spoken explanation. Clark (1997) has called this perspective the philosophy of the *extended mind*.

In the context of complexity, what counts is not whether we classify an information medium as "mind" or as "matter", but how we can ensure that this channel will pass on the right information at the right time. Classical thinking is completely unhelpful in this respect, since it assumes that in order to make rational decisions we simply need *all* information about the system under consideration, since otherwise there is indeterminacy. The complexity perspective notes not only that indeterminacy is unavoidable, but that having too much information is as bad as having too little. Indeed, we only need the difference that makes a difference *with respect to our goals*: what happens in the rest of the world is irrelevant, and trying to include it in our models will only burden the very limited capacities we have for information processing. In complex thinking, knowledge is a subjective model constructed to make problem-solving as easy as possible, not an objective reflection of outside reality, and decision-making aims for the solution that is good enough while using any opportunity to improve it later, rather than for the optimal one that classical thinking presupposes. The crux therefore is to be selective, and distil the essential *difference* from the infinite supply of data.

Happily, self-organization again comes to the rescue. An organization formed by a collection of agents connected by a variety of communication channels will spontaneously improve itself by *learning* from experience: communications that are effective will be used more often, while those that are ineffective will be increasingly ignored. In that way, information will be propagated ever more efficiently, filtering out the noise, while making sure the right messages reach the right places. This learning happens simultaneously on the two levels: "mentally"—when a person experiences that another person reacts either appropriately or inappropriately to a particular type of message and thus becomes more discriminative about whom to communicate with about what, and "materially"—when more useful supports, such as a handbook or Internet connection, are made more accessible, while inefficient ones, such as phone directory or fax machine, are relegated to the dust heap. However, this distributed learning is still pretty slow and unreliable in typical organizations. Applying the principles of self-organization and cybernetics (Heylighen, 1992, 2002; Heylighen & Joslyn, 2001) may help us to design a much more effective organization.

Most concretely, we can build this self-organized learning into the very media rather than have it be imposed on them by the users. For example, when employees are bombarded every day with email messages from colleagues and departments, it would be helpful to prioritize these automatically, so that the most important messages—such as a request for an urgent meeting with the CEO—are attended to immediately, while the less important ones—such as the latest offerings at the company canteen—are classified at the bottom of the heap, where they can be attended to if there is time, but otherwise safely ignored. This ordering by priority is something that the communication system could learn from experience, by observing which kinds of messages tend to be attended to most actively by which employees. A slightly more sophisticated version of the system would also learn the typical workflow and division of labor in an organization, making sure that the right task is sent to the right person at the right moment. Our research group is presently investigating through computer simulation how a group of agents using various communication media can self-organize so as to achieve such coordination (Crowston, 2003) between their activities .

## 8. Conclusion

We still do not understand complexity very well, and there is much to be done and explored in this direction. Our culture now is immersed and surrounded by complexity. But facing this complexity forces us to change our ways of thinking (Heylighen, 1991). We have shown how classical thinking, with its emphasis on analysis, predictability and objectivity, breaks down when confronted with complex systems. The core problem is that classical philosophy assumes invariant, conserved distinctions, whereas complex systems are entangled in such a way that their components and properties can no longer be separated or distinguished absolutely. Moreover, because of the inherent non-linearity or "loopiness" of the system, they tend to change in a chaotic, unpredictable way. At best, we can make context-dependent distinctions and use them to build a partial model, useful for a particular purpose. But such model will never be able to capture all essential properties of the system, and a novel context will in general require a different model.

This chapter did not so much propose specific tools and techniques for managing complex systems, as this would require a much more extensive discussion. Moreover, introductions to and reviews of existing concepts are available elsewhere (e.g. Kelly, 1994; Heylighen, 1997; Battram, 2002). Instead we have brought forth a number of ideas that allow us to better understand and speak about complex systems. First, we must be aware that complex systems are never completely predictable, even if we know how they function. We should be prepared to deal with the unexpected events that complexity most certainly will bring forth, by as quickly as possible correcting any deviation from our planned course of action. To achieve this kind of error-based regulation we should not try to predict or *determine* the behaviour of a complex system, but to *expect* the most probable possibilities. This will make it easier for us to *adapt* when things go off-course. Because then we are ready to expect the unexpected.

Yet, complexity does not just make our life harder. By better understanding the mechanism of self-organization, we may make a complex system work for us. This we could do by preparing the conditions so that self-organization is stimulated in a way that fits with our desires. We may even make use of, and stimulate, the inherent intelligence that characterizes a self-organizing system. By stimulating people as well as media to interact and learn from their interactions we may foster a distributed form of information-processing that coordinates the different activities in the system.